\documentclass[10pt]{article}

\usepackage{amsmath}
\usepackage{amssymb}
\usepackage{amsthm}
\usepackage{colonequals}
\usepackage{enumerate}

\newtheorem{definition}{Definition}
\newtheorem{fact}{Fact}
\newtheorem{remark}{Remark}
\DeclareMathOperator{\cpc}{{\mathit{cpc}}}
\DeclareMathOperator{\val}{{\mathit{rule}}}
\DeclareMathOperator{\throttle}{{\mathit{throttle}}}

\DeclareMathOperator{\impProfit}{\mathit{imp\_profit}}
\DeclareMathOperator{\score}{\mathit{score}}
\DeclareMathOperator{\ctr}{{\mathit{ctr}}}

\begin{document}

\title{Behavioral On-Line Advertising}

\author{
Fabrizio Caruso \\ Neodata Group, Catania \\ \texttt{fabrizio.caruso@neodatagroup.com} \and 
Giovanni Giuffrida \\ Dept. of Social Sciences, University of Catania \\ \texttt{ggiuffrida@dmi.unict.it} \and 
Calogero Zarba \\ Neodata Group, Catania \\ \texttt{calogero.zarba@neodatagroup.com}}

\maketitle

\begin{abstract}
We present a new algorithm for behavioral targeting of banner
advertisements. 
We record different user's actions such as
clicks, search queries and page views.
We use the collected information on the user to estimate in real time 
the probability of a click on a banner.
A banner is displayed if it either has the highest probability
of being clicked or if it is the one that generates the highest average profit.

\medskip

\end{abstract}

\section{Introduction}
The setting of our problem is the following: we are given a finite set
of users, and a webserver.  At each instant of time, a user $u$ may
connect to the webserver, requesting a webpage.  The webserver
responds to the request, and inserts into the webpage an appropriate
banner containing an advertisement.  The user may then click on the
banner, or he may not.

We take into account different events: impressions (visualizations), clicks,
registrations, page views, keywords in a search queries, etc\dots\ An
impression event occurs when the webserver responds to a user request
for a given webpage, and inserts into the webpage a banner.  A click
event occurs when the user clicks on a banner.  A registration is a
voluntary action of the user after a click on the banner such as a
purchase of the advertised item or the registration into a site and
may have different levels depending on the profit/value of the action
for the advertiser.  A page view is a simple view of a page.  A
keyword in a search query is the action of search for a specific word
in a search engine embedded in a website.  We refer to feature events
as to the events that can be used to study the behavior of the
users. We do not consider clicks and registrations as features
because in our model we are assuming the independence of features
(see equation~\ref{eq:Bayes} in Section~\ref{se:alg} for the technical details).
In Section~\ref{se:clicks} we describe a heuristic improvement of our method
that also considers clicks as features, similarly
to methods used in collaborative filtering. 
A good choice could be to take as features all voluntary feature events (except impressions).  

Each impression, click or registration (purchase of a product, registration
into the advertised site, etc\dots) of a banner $b$ can generate a profit.
For sake of simplicity we will primarily consider profits
generated by clicks and shortly give a description on how both
impressions and 
registrations (see Section~\ref{se:imps}) can be taken into consideration.

Our goal is to maximize the profit generated by all clicks or
maximize the number of clicks. The former goal is more general
than the latter, because if we assume unitary value for all clicks,
the total profit equals the total number of clicks.
This problem has already been treated in the scientific literature
(see \cite{ChenPavlovCanny}
where the linear Poisson regression model 
is used to predict click-through rates
and \cite{YanLiuWangZhangJiangChen} 
where a metric
is introduced to assess the quality of the behavioral targeting).
Our proposed approach is simpler than the other similar
approaches in the literature, in that it uses the naive Bayesian model.
Other approaches are possible such as the ones based on
linear programming models (\cite{abenakamura}, \cite{nakamura}, \cite{OptimizingOnLineAd}).

For a given user $u$, we store all events of the user
in the cookie maintained by
the browser.  The cookie contains also the timestamps of each
stored event.
The user's cookie is used by the webserver each time the user requests
a webpage, in order to select an appropriate banner.  

In this paper, we describe an algorithm that allows the webserver to select 
an appropriate banner, based on the information stored in the user's cookie.  

We refer to a {\em feature} as to the presence of a 
feature event in the user's cookie.
We try to estimate the value of the association rule
$f_1,\dots, f_n \to b$, i.e., the probability that a user $u$ clicks on $b$,
provided the $u$ has features $f_1,\dots,f_n$.

We use information on the features and the
click-through rates.
The webserver keeps track in real time of the
click-through rates of all banners among users that have the same
feature.

Moreover we propose a heuristics to avoid overflooding a user
with the same banner based on the impressions in the user's cookie.

The paper is organized as follows:
in Section~\ref{se:pre} we introduce some notation and definitions;
the algorithm for selecting
the banner to display is described in Section~\ref{se:alg};
the heuristics that limits the number of displays of the same banner
is described in Section~\ref{se:thr};
in the last sections different generalizations of the approach
are considered (impressions and registrations in Section~\ref{se:imps},
clicks as features in Section~\ref{se:clicks}, generalizations
in terms of locations and time are considered in Section~\ref{se:space-time}).

\section{Preliminaries}\label{se:pre}
We call ``impression'', the display of an advertisement. 
We will be using three data-structures to store information on
user's clicks and impressions at real time: 
a user's {\em history}, which depends on the user;
a {\em click matrix} and an {\em impression matrix}, which are global.

In the following we denote with $B=\{b_1,\dots, b_n\}$ the set of all banners
and by $\mathcal{F}$ the set of all features to be taken into
consideration.

\begin{definition}[User's history]
For every user $u$ we define their {\em history} 
as the set of triples $(\mathsf{type}, \mathsf{obj}, \mathsf{time})$
that describe all events and timestamps of user $u$,
where $\mathsf{type}$ is the type of event 
(impression, click, page view, search query, etc\dots),
$\mathsf{obj}$ is either the clicked banner,
the URL of the viewed page or the keyword in search query,
and where $\mathsf{time}$ is the timestamp of the event. 

\end{definition}

\begin{definition}[User's profile]
For every user $u$ we define their {\em profile} 
$\mathcal{P}_u=(\mathcal{F}_u,\mathcal{S}_u,\mathcal{C}_u)$,
where 
$\mathcal{F}_u \subseteq \mathcal{F}$,
$\mathcal{S}_u:B \to \mathbb{N}$
maps each banner to the number of its impressions to user $u$,
and 
$\mathcal{C}_u:B \to \mathbb{N}$ maps each banner to the number 
of its clicks by user $u$.
\end{definition}

\begin{remark}
The user's history is the only data-structure that needs to be
stored in the user's cookie.  
We have introduced the user's profile for the sake of simplicity.
\end{remark}

\begin{definition}
We denote by $S=(S_{i,j})$ the {\em impression matrix},
where $S_{i,j}$ is the number of
impressions of banner $b_j$ among users $u$ that have feature $i$,
i.e., $i \in \mathcal{F}_u$.
\end{definition}

\begin{definition}
We denote by $C=(c_{i,j})$ the {\em click matrix},
where $c_{i,j}$ is the number of clicks of banner $b_j$
among users $u$ who have feature $i$, i.e., 
$i \in \mathcal{F}_u$. 
\end{definition}

\section{The banner selection algorithm}
\label{se:alg}

Assume that a user $u$ requests a webpage from the webserver, which
responds by sending the webpage, and inserting into the webpage an
appropriate banner.  
We now describe the general strategy on how the banner is selected, based on 
the information stored in the user's history of $u$.
We will denote by $P(b)$ the global probability for a banner $b$
to be clicked and by $P(b \mid f_1, \dots, f_n)$ the
probability for a banner $b$ to be clicked by a user with
features $f_1, \dots, f_n$. We are also assuming that
$P(b) \neq 0$, $P(b \mid f_i) \neq 0$, $\forall i$.  
In particular we want to maximize the
probability $P(b \mid f_1, \ldots, f_n)$.

Given a set $K$ 
of candidate banners that may be
selected by the webserver,
for each $b \in K$ we compute a score $\score(b)$.
The banner with the highest score is then selected by the webserver.

As score we take
\begin{equation}
\score(b) = \cpc(b) \cdot \val(f_1, \dots, f_n \to b);
\end{equation}
\noindent where
\begin{itemize}
\item $\cpc(b)$ is the ``cost per click'' (the profit) of $b$;
\item $\val(f_1,\dots, f_n \to b)$ is defined as follows:
\begin{equation}
\val(f_1,\dots, f_n \to b) = P(b) \prod_{i=1}^n \frac{P(b \mid f_i)}{P(b)}.
\end{equation}
\end{itemize}

Given two expressions $\alpha$ and $\beta$ we use
use the notation $\alpha \propto \beta$ to mean 
$\alpha = c \beta$ where $c$ does not depend on $b$.

In particular we have the following fact

\begin{fact}
Under the hypothesis that the features $f_i$ are independent
events we have 
\begin{equation}
\val(f_1, \dots, f_n \to b) \propto P(b \mid f_1, \dots, f_n).
\end{equation}
\end{fact}
\begin{proof}
By applying Bayes' Theorem twice, under the simplifying
hypothesis of independent features, we have
\begin{equation}
\label{eq:Bayes}
\begin{split}
P(b \mid f_1, \ldots, f_n) &= 
\frac{P(b) P(f_1, \ldots f_n \mid b)}{P(f_1, \ldots, f_n)} \; \propto \; P(b) P(f_1, \ldots f_n \mid b) \\ 
&= P(b) \prod_{i=1}^{n} P(f_i \mid b) = P(b) \prod_{i=1}^{n} \frac{P(b \mid f_i) P(f_i)}{P(b)} \\
  &= P(b) \prod_{i=1}^n \frac{P(b \mid f_i)}{P(b)} \prod_{i=1}^n P(f_i) 
\; \propto \; P(b) \prod_{i=1}^n \frac{P(b \mid f_i)}{P(b)} \\
 &= \val(f_1, \dots, f_n \to b)\,.
\end{split}
\end{equation}
\end{proof}

Therefore the banner with highest $\val$ is the banner with the
highest probability of been clicked and the banner with
highest score is the banner that generates the highest average
profit per click.

\subsection{Click-through rates}\label{se:ctr}

In order to compute the probabilities 
$P(b)$, $P(b \mid f_i)$ for $i=1,\dots, b$.
in~\eqref{eq:Bayes} we use the concept of click-through rate:
\begin{equation}
\begin{split}
\ctr(b) &= 
\frac{\#\text{clicks on $b$}}{\#\text{impressions of $b$}}
\; (\text{among all users}). \\
\ctr_{f}(b) &= 
\frac{\#\text{clicks on $b$}}{\#\text{impressions of $b$}}
\; (\text{among users with feature $f$}).
\end{split}
\end{equation}

Therefore we can compute the probabilities $P(b)$, $P(b \mid f_i)$
as click frequencies.
Hence we have
\begin{equation}
P(b) = \ctr(b); \; \;
P(b \mid f_i) = \ctr_{f_i}(b).
\end{equation}

Thus we can write
\begin{equation}
\val(f_1, \dots, f_n \to b) = 
\ctr(b) \cdot \prod_{i=1}^n \frac{\ctr_{f_i}(b)}{\ctr(b)} 
= \ctr(b)^{1-n} \prod_{i=1}^n \ctr_{f_i}(b).
\end{equation}

\subsection{Updating relative click-through rates}\label{sse:upd}
In order to keep up to date the click-through rates for each feature
we need to update the impression matrix and the click matrix
in real time.
A click-through rate of banner $b_j$ for a given feature $i$ 
is then computed by counting the clicks and impressions in the $i$-th rows
of the two matrices:
\begin{equation}
\ctr_i(b_j) = \frac{c_{i,j}}{s_{i,j}}.
\end{equation}

We consider the set $B=\{b_1, \dots, b_n\}$ of all banners.
In order to update the matrices after each impression and click, 
for every user, first, the user's history (and profile)
are updated,
and second, the following actions on the matrices are taken:
\begin{itemize}
\item If there is an impression on banner $b_j$ by user $u$:
then
for every feature $i$ of $u$, 
$s_{i,j}$ is increased by one:
for every $i$ such that $i \in \mathcal{F}_u$ 
we do $s_{i,j}\colonequals s_{i,j}+1$.
\item If there is a click on banner $b_j$ by a user $u$ 
that has already clicked on $b_j$,
i.e.\ $\mathcal{C}_u(b_j)>0$:
then for every feature $i$ of $u$, 
$c_{i,j}$ is increased by one, i.e.\ 
for every $i$ such that $i \in \mathcal{F}_u$ 
we do $c_{i,j}\colonequals c_{i,j}+1$.
\item If there is a feature event $i$ by a user $u$ that
did not have that feature:
the $i-th$ rows in $C$ and $S$ are updated
with respectively the impressions and the clicks in the user's profile:
$s_{i,j}\colonequals s_{i,j}+\mathcal{S}_u(b_j)$,
$c_{i,j}\colonequals c_{i,j}+\mathcal{C}_u(b_j)$, $\forall j$. 
\end{itemize}

\section{Avoiding user's boredom}\label{se:thr}
In this section we describe our strategy on how to avoid overflooding
a user with the same banner.
We achieve this by ``throttling down'' the value of the candidate banner
taking into account the times at which the banner has already been
displayed to the user.
The value $\val(f_1,\dots, f_n \to b)$ is multiplied
by a scaling factor $\throttle_u(b)$
with the following properties: 
\begin{enumerate}[I.]
\item $0<\throttle(b)\le 1$: in order to have a down-scaling.
\item $\throttle(b)$ decreases with the number of impressions of~$b$.
\item $\throttle(b)$ decreases more if the impressions are more recent
and it increases if the impressions are farther in the past.
\end{enumerate}

So that we have 
\begin{equation*}
  \score(b) = 
  \cpc(b)
  \val( f_1,\dots, f_n \to b) 
  \throttle(b) \,.
\end{equation*}

In particular we choose the following function.
Let $t$ be the current instant of time.  Also, let $t_i$, for $i =
1, \ldots, m$ be the instants of time in which an impression event for
banner $b$ and user $u$ has occurred. 

\begin{equation}\label{eq:throttle}
  \throttle(b) = \prod_{i=1}^m \left( 1 - \alpha
  \left(\frac{1}{2}\right)^{(t - t_i) / h}\right) \,,
\end{equation}
\noindent where $0<\alpha<1$ and $h$ are heuristically selected parameters.

Thus the formula in (\ref{eq:throttle})
can avoid overflooding the user with the same banners
and can improve the estimation of the probability of a click.
Therefore we have the following facts
\begin{itemize}
\item $\val(f_1, \dots, f_n \to b) \throttle(b)$ 
is an approximation of the probability of a click on banner
$b$ by a user who has features $f_1, \dots, f_n$ and has already seen certain
impressions of banner $b$. 
\item $\score(b)$ is an approximation of the expected average profit 
solely generated
by a possible click
after an impression of banner $b$.
\end{itemize}

\section{Impressions and Registrations}\label{se:imps}
In the most general case we may have banners that 
generate a profit for each impression, click 
and registration.

For each candidate banner $b$, 
we can take into account profits generated by
both impressions and clicks by considering:
\begin{equation}
\score^{+}(b) = \impProfit(b) + score(b);
\end{equation}
\noindent where $\impProfit(b)$ is the profit 
generated by the display of $b$.

This approach can be further generalized to encompass
registrations of any step by simply treating them as 
click-like events.

\section{Click events as features}\label{se:clicks}
Our approach only considers non-click events
as features and assumes that the features
are independent of each other.
We can improve the accuracy of our predictions by considering
click events but they would have to be
treated differently because the basic
assumption of independence does not hold for them.
Clicks could be treated similarly to a purchase
in a collaborative filtering approach.
We record each unique click in a 
$\text{user} \times \text{banner}$ matrix.
The probability $P(b \mid c)$ of a click
on banner $b$ by a user that has clicked
on $c$ will then depend on whether others
users that have clicked on $c$ have 
also clicked on $b$.
For more details on an practical use
of this approach we refer to
\cite{UniqueCollFilt}, where the
approach is used for a recommending system.

\section{Space and time}\label{se:space-time}
This approach can be generalized
in terms of time and space i.e.,
location of a banner.

We can also take into account these significant
attributes
in order to better target the users at specific
times for banners at specific locations.
This can be achieved by simply recording this
data in an extra dimension in the matrices $S$, $C$ described
in Section~\ref{sse:upd}.

\section{Future work}
This approach could be further developed, improved
and generalized in different respects:
with respect to how the features are treated
(by considering clusters of features instead of
single features or by considering
non-Boolean features), and with respect
to its applications.

\subsection{Non-Boolean features}
We can assign each feature a counter that could
be used in an extended definition of value of $\val(f_1,\dots, f_n \to b)$.
One possibility could be to consider
\begin{equation}
\val^{*}(f_1,\dots, f_n \to b) :=
W(c_1,\dots, c_n) \cdot \val(f_1,\dots, f_n \to b),
\end{equation}
\noindent where 
\begin{itemize}
\item $c_i$ is the (possibly normalized with respect with the average)
counter of feature $f_i$, for $i=1,\dots, n$;
\item $W(c_1,\dots, c_n)$ is a measure of how the
counters should correct the association rule $f_1,\dots, f_n \to b$.

A possible straightforward candidate for $W(c_1,\dots, c_n)$
could be the simple arithmetic average:
\begin{equation}
W(c_1,\dots, c_n) := \frac{c_1 + \dots + c_n}{n}.
\end{equation}
\end{itemize}

This could be used to
differentiate between a single page view 
(or a single search query with a keyword) and multiple 
page views (multiple search queries with the same keyword).

\subsection{Application to on-line newspapers and magazines}
This approach could also be applied to on-line newspapers
and magazines in that the visualization of 
an article's title is seen as an impression and
a click on the title is seen as a click on a banner.

\bibliographystyle{plain}

\end{document}